\documentclass{aa}

\usepackage{multirow}
\usepackage{comment}
\usepackage{txfonts}
\usepackage{hyperref}
\hypersetup{colorlinks=True,citecolor=blue,allcolors=blue}

\usepackage{xcolor}

\begin{document}

\title{Microlensing analysis of 14.5-year light curves in SDSS~J1004+4112:
Quasar accretion disk size and intracluster stellar mass fraction}

\author{
R. For\'es-Toribio\inst{1,2}\orcid{0000-0002-6482-2180}
\and J. A. Mu\~noz\inst{1,2}\orcid{0000-0001-9833-2959}
\and C. Fian\inst{1,2}\orcid{0000-0002-2306-9372}
\and J. Jim\'enez-Vicente\inst{3,4}\orcid{0000-0001-7798-3453}
\and E. Mediavilla\inst{5,6}\orcid{0000-0003-1989-6292}
}

\institute{Departamento de Astronom\'{\i}a y Astrof\'{\i}sica, Universidad de Valencia, E-46100 Burjassot, Valencia, Spain
\email{raquel.fores@uv.es}
\and Observatorio Astron\'omico, Universidad de Valencia, E-46980 Paterna, Valencia, Spain 
\and Departamento de F\'{\i}sica Te\'orica y del Cosmos, Universidad de Granada, Campus de Fuentenueva, 18071 Granada, Spain
\and Instituto Carlos I de F\'{\i}sica Te\'orica y Computacional, Universidad de Granada, 18071 Granada, Spain
\and Instituto de Astrof\'{\i}sica de Canarias, V\'{\i}a L\'actea S/N, La Laguna, E-38200, Tenerife, Spain 
\and Departamento de Astrof\'{\i}sica, Universidad de la Laguna, La Laguna, E-38200, Tenerife, Spain
}

\date{Received xx / Accepted yy}

\abstract{The gravitational lens system SDSS~J1004+4112 was the first known example of a quasar lensed by a galaxy cluster. The interest in this system has been renewed following the publication of r-band light curves spanning 14.5 years and the determination of the time delays between the four brightest quasar images.}
{We constrained the quasar accretion disk size and the fraction of the lens mass in stars using the signature of microlensing in the quasar image light curves.}
{We built the six possible histograms of microlensing magnitude differences between the four quasar images and compared them with simulated model histograms, using a $\chi^2$ test to infer the model parameters.}
{We infer a quasar disk half-light radius of $R_{1/2}=(0.70\pm0.04)\, R_E=(6.4\pm0.4) \sqrt{M/0.3M_{\sun}}$ light-days at 2407\AA\ in the rest frame and stellar mass fractions at the quasar image positions of $\alpha_A>0.059$, $\alpha_B=0.056^{+0.021}_{-0.027}$, $\alpha_C=0.030^{+0.031}_{-0.021}$, and $\alpha_D=0.072^{+0.034}_{-0.016}$.}
{The inferred disk size is broadly compatible with most previous estimates, and the stellar mass fractions are within the expected ranges for galaxy clusters. In the region where image C lies, the stellar mass fraction is compatible with a stellar contribution from the brightest cluster galaxy, galaxy cluster members, and intracluster light, but the values at images B, D, and especially A are slightly larger, possibly suggesting the presence of extra stellar components.}


\titlerunning{Microlensing of SDSS~J1004+4112:
Accretion disk size and stellar mass fraction}
\authorrunning{For\'es-Toribio et al.}

\maketitle

\section{Introduction} \label{sec:intro}

Since \citet{1979Natur.282..561C} proposed that stars located near the light paths of gravitationally lensed quasars may cause flux variations in the quasar images, the field of gravitational microlensing has been developed and refined such that now we are able to estimate the sizes of quasar accretion disks \citep[see e.g.][]{2004ApJ...605...58K,2012ApJ...751..106J,2013ApJ...774...69H,2015ApJ...798...95B,2015ApJ...806..258M,2016ApJ...817..155M,2018ApJ...869..106M,2021A&A...654A..70F} and the abundance of microlenses in the lens \citep[see e.g.][]{2009ApJ...706.1451M,2015ApJ...799..149J,2023A&A...673A..88A}.

Quasars also exhibit intrinsic flux variability that needs to be removed in order to obtain the fluctuations produced only by microlensing. If the time delay between images is known, we can subtract pairs of light curves shifted by their time delays to obtain light curves of the microlensing magnification differences. We also need sufficiently long light curves to statistically sample the magnifications and infer the system parameters properly. With the recent measurements of 14.5-year light curves and time delays for the four brightest images of SDSS~J1004+4112 \citep{2022ApJ...937...34M}, we are now able to carry out such studies.

SDSS~J1004+4112 was the first discovered lens system in which a background quasar is lensed by a galaxy cluster instead of a single galaxy \citep{2003Natur.426..810I}. This leads to a large 15\arcsec\ separation between the images, and hence the light from the quasar travels mainly through the cluster instead of passing through an individual galaxy, as is typical for lensed quasars. With this configuration, the impact of microlensing in the quasar was expected to be small. However, shortly after its discovery, \citet{2004ApJ...610..679R} reported microlensing variability in the blue wing of image A in several emission lines corresponding to the broad line region, and this variability continued over the years \citep{2006ApJ...645L...5G,2006A&A...454..493L,2012ApJ...755...82M,2018ApJ...859...50F,2020A&A...634A..27P,2021A&A...653A.109F}. The main explanation for these fluctuations is microlensing, but other possibilities have been raised, such as outflows \citep{2006ApJ...644..733G,2007ApJ...658..748A,2020A&A...634A..27P}, because the stellar density in galaxy clusters is expected to be insufficient to drive such microlensing variability.

Microlensing variability in the continuum emission of the accretion disk is also observed \citep{2008ApJ...676..761F,2012ApJ...755...24C,2016ApJ...830..149F}. \citet{2008ApJ...676..761F} and \citet{2016ApJ...830..149F} used the light curves to determine the quasar disk size in the r band. However, \citet{2008ApJ...676..761F} used only images A and B, while \citet{2016ApJ...830..149F} used all four images but with the predicted time delay for image D from \citet{2010PASJ...62.1017O}, which now we know from \citet{2022ApJ...937...34M} was 413 days too short.

We used the newly measured time delays along with the light curves from \citet{2022ApJ...937...34M} to constrain the quasar disk size, and, given the quality and long coverage of the light curves, we simultaneously attempted to measure the fraction of the mass in stars in the regions where the quasar images are located. The paper is structured as follows. In Sect. \ref{sec:data} we present the data that we used for the analysis, Sect. \ref{sec:method} describes the methodology, the results are presented in Sect. \ref{sec:results}, and we discuss and summarise them in Sect. \ref{sec:disc}.

\section{Data} \label{sec:data}

We used the r-band light curves of the four brightest quasar images from \citet{2022ApJ...937...34M}, which span 14.5 years. These light curves were shifted by the time delays $\Delta t_{AC}=825.99\pm0.42$, $\Delta t_{BC}=781.92\pm0.44,$ and $\Delta t_{DC}=2456.99\pm1.11$ days, also from \citet{2022ApJ...937...34M}. We assumed that the weighted means of the infrared (IR) 8.0 $\mu m$ joint quasar and host (QSO+Host) magnitude measurements of \citet{2009ApJ...702..472R} are a solid proxy for the un-microlensed flux ratios: $m_{A,IR}=12.262\pm0.009$, $m_{B,IR}=12.627\pm0.015$, $m_{C,IR}=12.906\pm0.007,$ and $m_{D,IR}=13.483\pm0.013$. They are compatible with radio flux ratios \citep{2020ApJ...900L..15J,2021MNRAS.505L..36M,2021MNRAS.508.4625H} and with the broad line region cores of \citet{2024A&A...682A..57F}. Finally, we used the values of convergence and shear at the image positions from the mass model of \citet{2022ApJ...937...35F} to simulate the magnification patterns.

\section{Methodology} \label{sec:method}

\subsection{Observed histograms} \label{sec:method-obs}

To reduce the noise, we smoothed the light curves over a window of 10 days. The smoothing window is much smaller than the estimated microlensing source-crossing timescale of $t_S=0.28$ yr \citep{2011ApJ...738...96M}. The average of the data points inside the window was weighted by their uncertainties. 

\begin{figure}
    \centering
    \includegraphics[width=\linewidth]{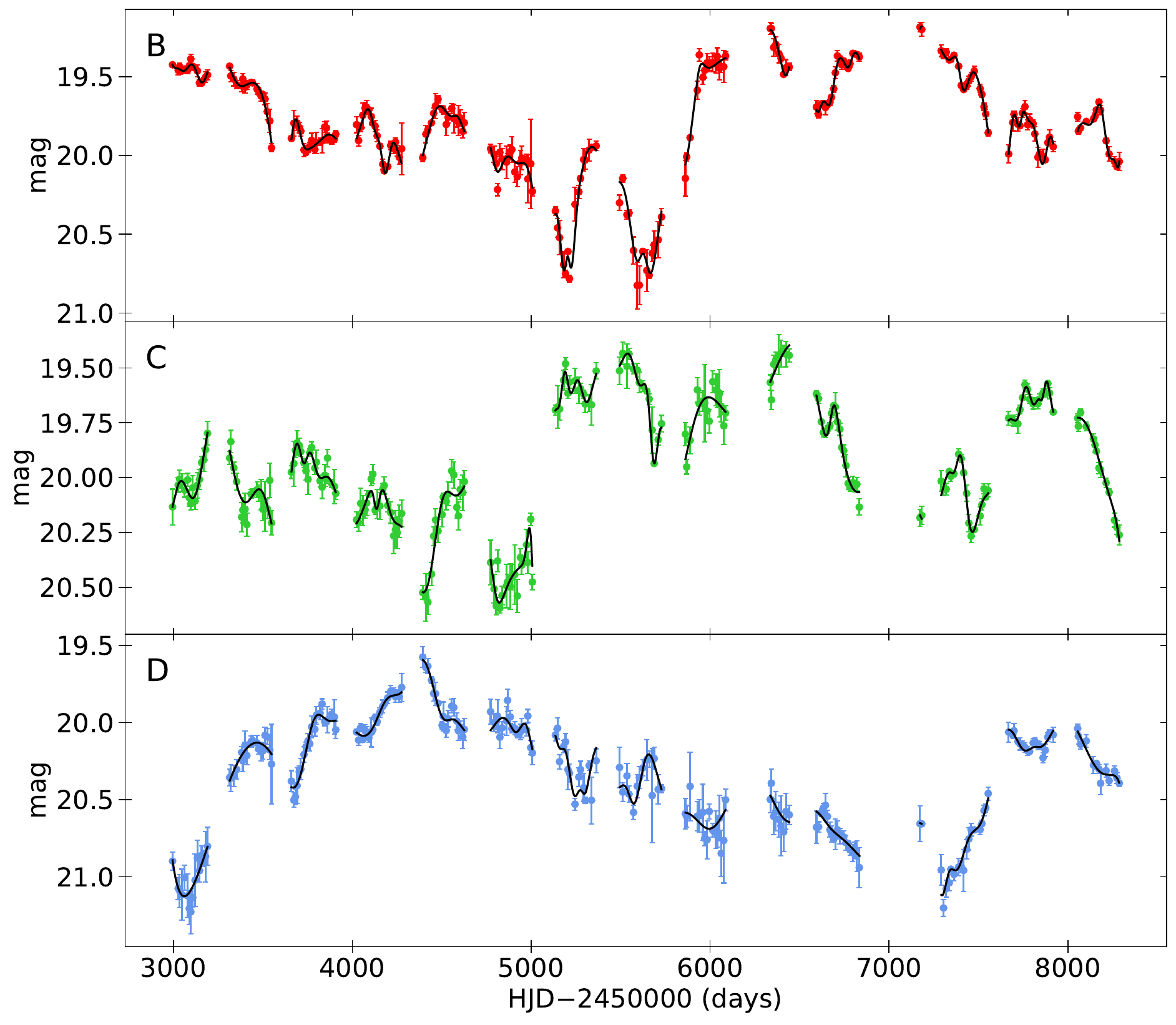}
    \caption{Fits to the r-band light curves B, C, and D from \citet{2022ApJ...937...34M} using fifth-order splines (solid black curves). Since gaps longer than 100 days were not considered in the analysis, the splines in these regions are not displayed. The days on the x-axis correspond to the observation dates.}
    \label{fig:spline-fits}
\end{figure}

The aim was to remove the intrinsic variability of the quasar, so we subtracted the magnitudes of one light curve from those of another after shifting them by their corresponding time delays. In order to do so, we needed to fit one of the light curves to interpolate between points. We computed the six possible differences --- A$-$B, C$-$B, D$-$B, A$-$C, D$-$C, and A$-$D --- and we chose to fit the light curve that was going to be subtracted. We fitted this light curve with a fifth-order spline, which gives enough flexibility to account for the fluctuations in the light curves without introducing extra structure. We determined the fit to be unreliable if the gap between two points is larger than 100 days, and we also manually removed the peaks around 5200 and 5600 days (heliocentric Julian date of observation minus 2450000) in the light curve of B, which give residuals that can be associated with intrinsic variability. We show the spline fits of light curves B, C, and D in Fig. \ref{fig:spline-fits}.

\begin{figure*}
    \centering
    \includegraphics[width=\linewidth]{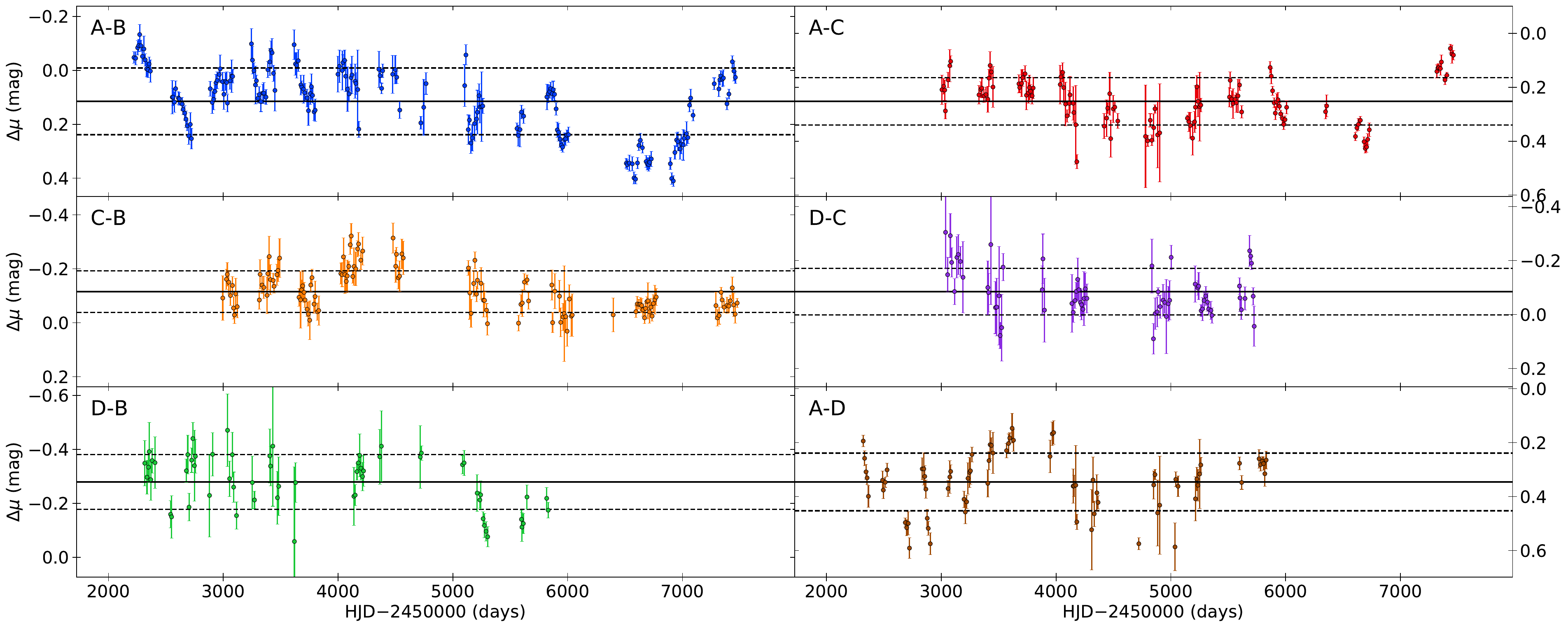}
    \caption{Microlensing differences between the six possible combinations of image pairs. The solid black lines are the mean of the micro-magnification, and the dashed lines correspond to the standard deviation. All light curves are shifted to match the time reference of the leading image, C.}
    \label{fig:micro-diff}
\end{figure*}

\begin{figure}
    \centering
    \includegraphics[width=\linewidth]{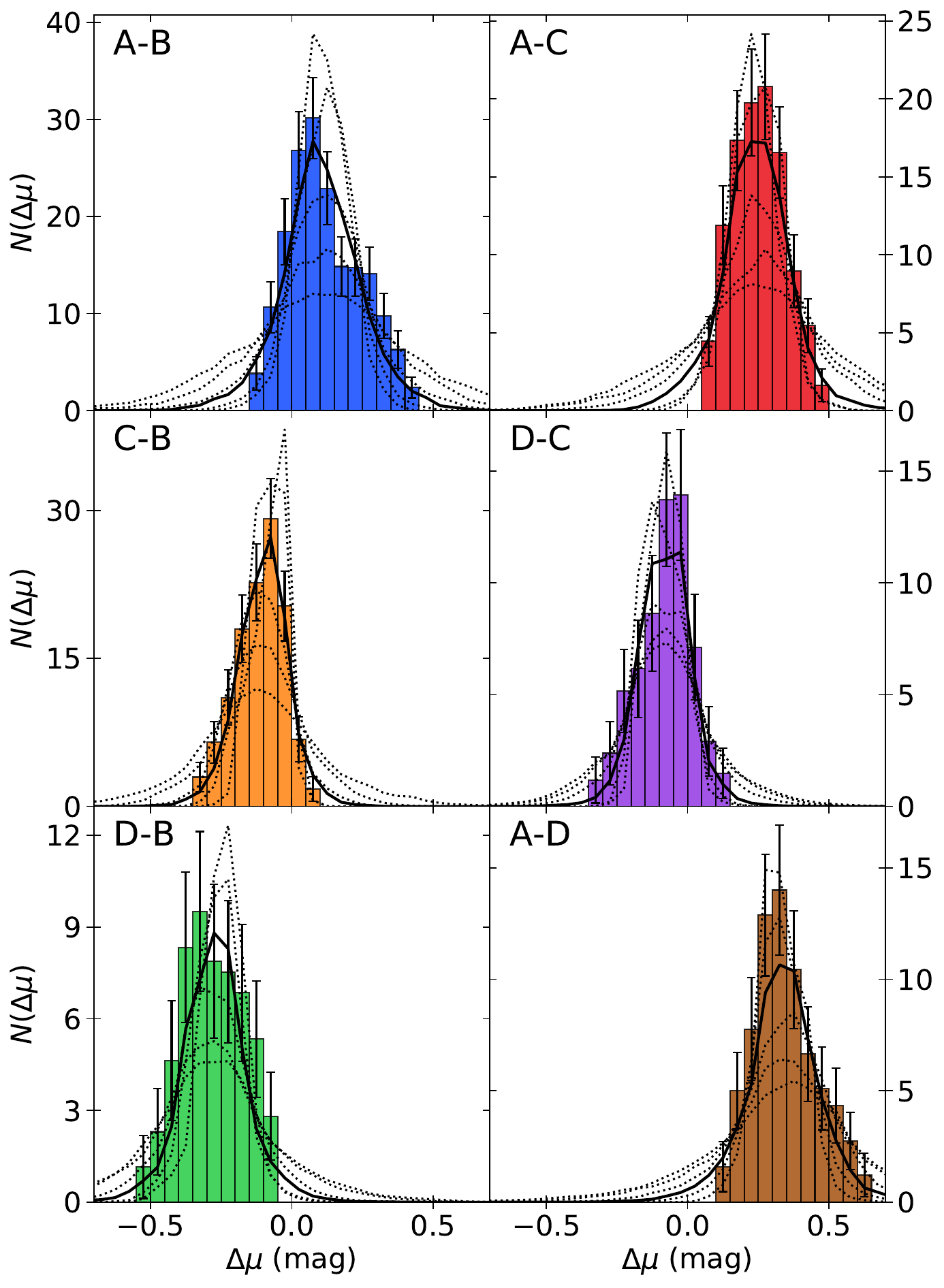}
    \caption{Observed histograms of the six possible microlensing differences obtained after averaging the $10^6$ realisations from the Monte Carlo sampling.\ The error bars are the standard deviation of these realisations. The solid black lines represent the best model difference histograms, and the thin dotted lines are other model histograms used for the parameter inference.}
    \label{fig:obs-hist}
\end{figure}

The differences between the shifted pairs of light curves still included the local macro-magnification of the lens system, which had to be removed. We took the 8.0 $\mu m$ measurements of \citet{2009ApJ...702..472R} as the macro-magnification baseline assuming that the emitting region of QSO+Host in this band is large enough to not be affected by microlensing. We computed the microlensing differences as
\begin{equation}
    \Delta \mu_{XR}^i= (m_{X}^i-m_{X,IR})-(m_{R,fit}^i-m_{R,IR}),
\end{equation}
where $m_{X}^i$ is an epoch of the X light curve that overlaps with the permitted regions of the R (reference) light curve, $m_{R,fit}^i$ is the value of the fitted reference light curve shifted by the corresponding time delay between images, and $m_{X,IR}$ and $m_{R,IR}$ are the IR magnitudes. The six possible microlensing differences generated by this procedure are shown in Fig. \ref{fig:micro-diff} with the corresponding mean and standard deviation. The light curve with the largest variability is A$-$B, as expected given the previous detections of microlensing in image A. In the A$-$B and A$-$C light curves, a clear microlensing event can be seen between days 6500 and 7500.

We constructed the observed histograms, $h_{XR}$, from these light curves taking the observational errors of each point into account  via Monte Carlo sampling, assuming that the errors of the data points are Gaussian. We generated $10^6$ histograms for each microlensing difference light curve. The histograms have a bin width of 0.05 mag and range from $-$5 to 5 mag. The value of each bin is the mean of the counts falling into each bin, and the associated error is the standard deviation; these histograms are shown in Fig. \ref{fig:obs-hist}.

\begin{figure*}
    \sidecaption
    \includegraphics[width=12cm]{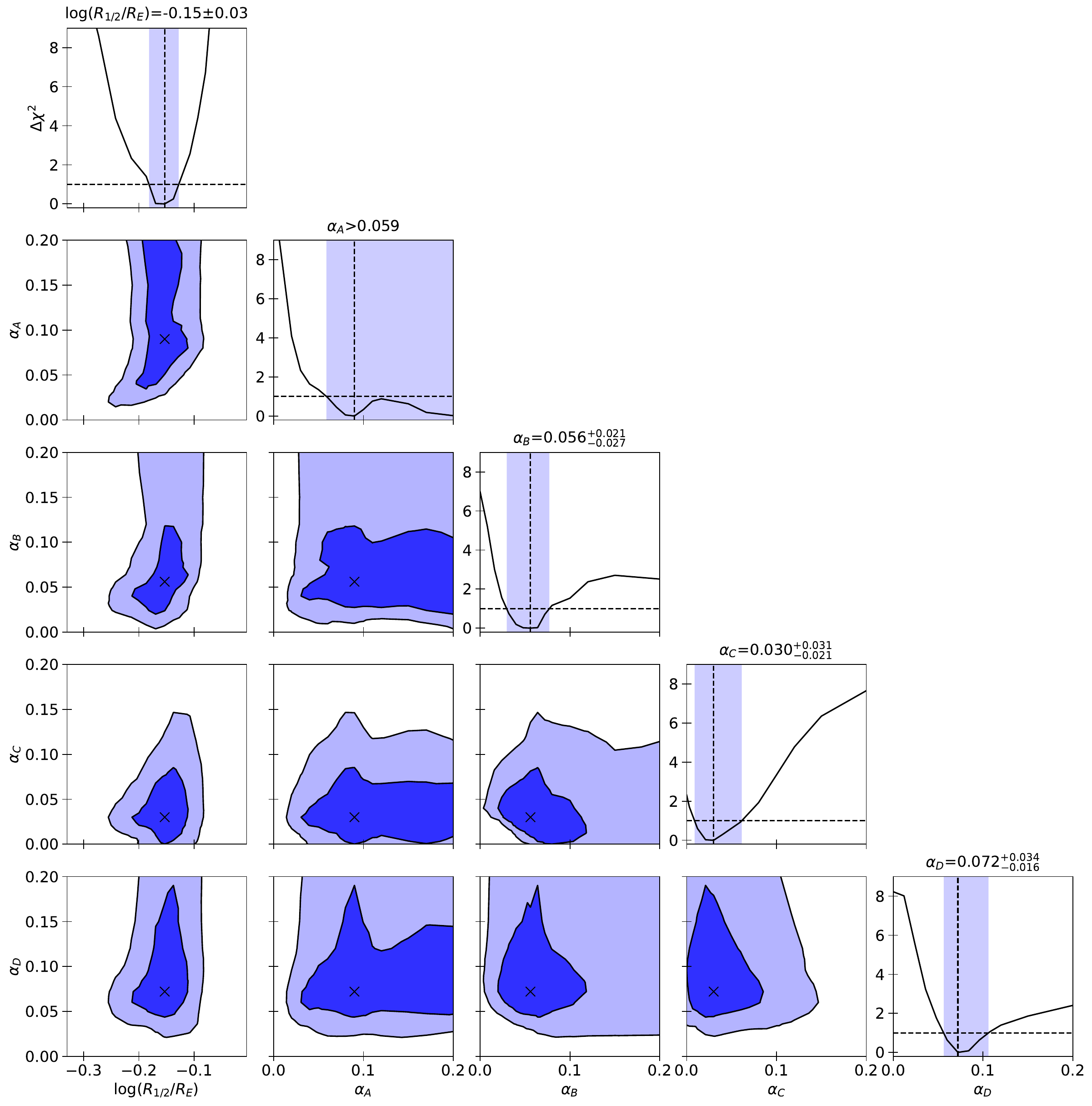}
    \caption{Joint $\Delta\chi^2=\chi^2-\chi^2_{\text{min}}$ contours for all pairs of parameters ($R_{1/2}$, $\alpha_A$, $\alpha_B$, $\alpha_C$, and $\alpha_D$). In the 2D likelihoods, the contours are drawn at the 68\% ($\Delta\chi^2=2.30$) and 95\% ($\Delta\chi^2=6.18$) confidence limits for two parameters. In the 1D distributions, the shaded regions are the 68\% ($\Delta\chi^2=1.00$) confidence regions for one parameter. The maximum likelihood model is marked with a cross and a dashed line for the 2D and 1D plots, respectively.}
    \label{fig:corner5}
\end{figure*}

\subsection{Model histograms} \label{sec:method-mod}

We computed magnification maps for the four images using the fast multipole method–inverse polygon mapping\footnote{https://gloton.ugr.es/microlensing/} developed by \citet{2022ApJ...941...80J}. We created 30$\times$30 $R_E$ magnification maps using stars of a constant mass of 0.3$M_{\sun}$. For this mass, the Einstein radius in the source plane corresponds to 9.07 light-days. Assuming an effective source velocity of 0.106 $R_E$/year \citep{2011ApJ...738...96M}, the magnification map resolution was set to $1$ pix $=0.003\,R_E$ so that one pixel covers the smoothing window of 10 days applied to the observed light curves. We computed the magnification maps with the values of $\kappa$ and $\gamma$ reported in \citet{2022ApJ...937...35F} and a variable fraction of the mass in stars, $\alpha=\kappa_{\ast}/\kappa$.

Once the magnification maps were created, we convolved them with extended sources of different sizes. We modelled the sources as a Gaussian of width $r_s$ since it is commonly accepted that the microlensing variability depends mainly on the half-light radius of the source rather than the particular source shape \citep{2005ApJ...628..594M,2016ApJ...817..155M,2019MNRAS.486.1944V}. After the convolution, the magnification maps were normalised in flux to remove the macro-magnification, as also done for the data.

According to \citet{2011ApJ...738...96M}, the effective velocity on the source plane is dominated by the random velocity of the microlenses in the galaxy cluster (517 km/s) rather than by the peculiar velocities of the source, the lens, and the observer (82 km/s, 196 km/s, and 99 km/s, respectively). Hence, the orientation of the paths that the source describes over the magnifications maps is essentially uncorrelated between image pairs \citep[see the discussion in][]{2010ApJ...712..658P}. Given the lack of significant correlations, we generated randomly oriented straight tracks with the same length as the observed difference light curves for each image pair on its corresponding magnification map. Since the observed histograms may arise from regions with different mean magnifications, we only kept the pairs of tracks whose average after subtracting one from the other differs by less than 0.1 mag from the mean micro-magnification difference of the observed histograms. This 0.1 mag is the error assumed for the un-microlensed flux ratios given the average error budget in IR found by \citet{2009ApJ...702..472R} and in broad emission line cores by \citet{2024A&A...682A..57F} for image pairs. We collected 500 pairs of tracks that fulfil this condition (see Fig. \ref{fig:tracks} as an example for the A$-$B pair) and obtained their microlensing difference histograms with the same bounds and widths as the observed histograms\footnote{These computations were performed at the computing service PROTEUS (\url{https://proteus.ugr.es}) of the Carlos I Institute of Theoretical and Computational Physics because of the large computational time required.}. Then, a single model histogram for each combination of source size and stellar mass fraction was computed by averaging the histogram bins of the 500 selected pair tracks to obtain
\begin{equation}
    \tilde{h}_{XR}(i;r_s,\alpha_X,\alpha_R)=\frac{1}{N}\sum_{n=1}^{N}\tilde{h}_{XR}^n(i;r_s,\alpha_X,\alpha_R),
\end{equation}
where $\tilde{h}_{XR}^n(i;r_s,\alpha_X,\alpha_R)$ is the $i$th bin of the histogram obtained from the $n$th selected track pair of the difference X$-$R for a source size $r_s$ and stellar mass fractions $\alpha_X$ and $\alpha_R$. The term $\tilde{h}_{XR}(i;r_s,\alpha_X,\alpha_R)$ is the $i$th bin of the model microlensing difference histogram for these parameters, which is the average of $N$=500 histograms.

\subsection{Statistical test}
\label{sec:method-test}

For each pair of images, we compared the observed microlensing histograms with the model histograms that arise from the possible combinations of source sizes, $r_s$, and stellar mass fractions, $\alpha,$ after normalising their counts. The likelihood that a given model with $r_s$, $\alpha_A$, $\alpha_B$, $\alpha_C$, and $\alpha_D$ reproduces the observed microlensing differences was computed using the $\chi^2$ test defined as
\begin{equation}
    \chi^2=\sum_{\mu}\sum_i \left(\frac{h_{\mu}(i)- \tilde{h}_{\mu}(i;r_s,\alpha_X,\alpha_R)}{\epsilon_{\mu}(i)}\right)^2.
\end{equation}
The sum in $i$ runs over the histogram bins, and the sum in $\mu$ runs over the six distinct image pairs (i.e. the difference between image X and the image of reference, R). The $h_{\mu}(i)$ is the $i$th bin of the normalised observed histogram built as described in Sect. \ref{sec:method-obs}, $\epsilon_{\mu}(i)$ is the error associated with that bin, and $\tilde{h}_{\mu}(i;r_s,\alpha_X,\alpha_R)$ is $i$th bin of the model histogram for the given set of parameters computed according to the procedure in Sect. \ref{sec:method-mod}. Given that low-count bins can affect the $\chi^2$ statistic, we only considered bins of the observed histograms with at least five counts \citep[see e.g.][]{2006smep.book.....J}.

\section{Results} \label{sec:results}

We computed the $\chi^2$ test for all the parameter combinations within the following ranges: We varied the source size, $r_s$, from 3.2 light-days to 7.6 light-days using logarithmically spaced values and then added more points near the $\chi^2$ minimum. The stellar mass fractions at the locations of images A, B, C, and D were varied in a range from 0 to 0.2. This range was set to not exceed the average stellar mass fraction found for individual galaxies \citep[e.g.][]{2015ApJ...799..149J}. Figure \ref{fig:corner5} shows the 68\% and 95\% maximum likelihood contours for two parameters ($\Delta \chi^2=2.30$ and $6.18$) and the probability distributions for each variable with their 68\% confidence limits ($\Delta \chi^2=1$). In this figure, a mild Gaussian smoothing is applied to obtain cleaner contours. The maximum likelihood histograms fit the observational data with a reduced $\chi^2$ of 0.997 (see the solid black lines in Figure \ref{fig:obs-hist}). The radius is presented as the logarithm of the half-light radius in Einstein radii, which is related to the Gaussian width as $R_{1/2}=1.18r_s$.

\begin{figure*}
    \centering
    \includegraphics[width=\linewidth]{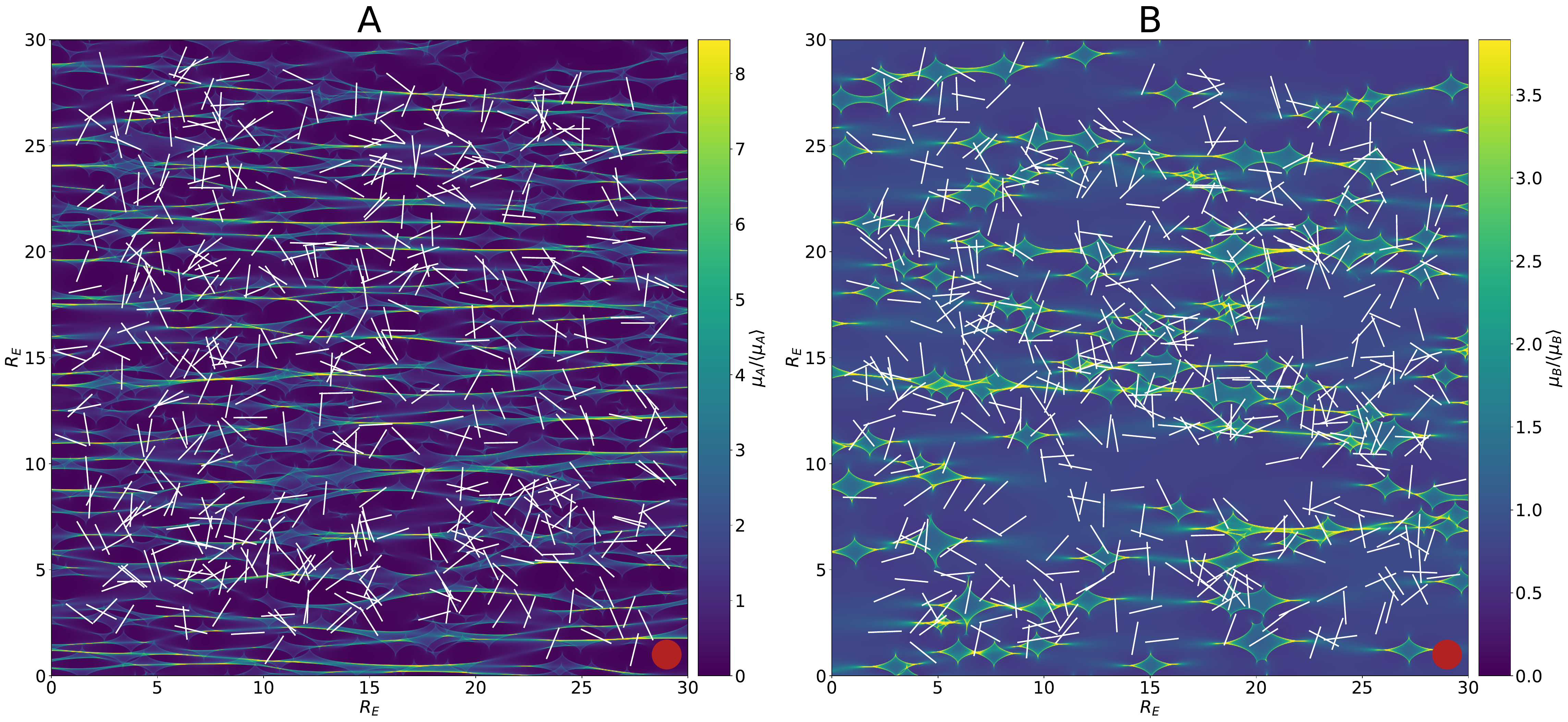}
    \caption{Magnification maps normalised to the mean macro-magnification, $\langle \mu \rangle$, for images A and B with the 500 pairs of tracks of the best fitting model. These tracks are randomly oriented, and their microlensing difference histograms were averaged to build the model histogram. The radius of the red circle at the bottom of each map corresponds to the inferred half-light radius of the source (i.e. 0.7$R_E$ or equivalently 6.4 light-days).}
    \label{fig:tracks}
\end{figure*}

With this procedure, we obtain an accretion disk size of $R_{1/2}=(0.70\pm0.04)\, R_E=(6.4\pm0.4) \sqrt{M/0.3M_{\sun}}$ light-days at 2407\AA\ in the rest frame. In Table \ref{tab:stars} we summarise the stellar mass fractions and the convergence in stars at the four quasar positions. For $\alpha_A$, we can only report a lower limit because of the poor convergence of the likelihood towards larger values, although these 
large values are unphysical. The magnification maps of images A and B along with the selected 500 pairs of tracks for the maximum likelihood model are shown in Fig. \ref{fig:tracks} as an example of the method.

\section{Discussion and conclusions} \label{sec:disc}

\begin{table}
\caption{Inferred stellar mass fraction and convergence in stars at the locations of the four quasar images.}
\label{tab:stars}
\centering
\begin{tabular}{c c c c}
\hline\hline
Image & $\alpha$ & $\kappa_{\ast}$ & d (kpc/h) \\
\hline
    A & $>0.059$ & $>0.042$ & $42.1\pm0.2$ \\
    B & $0.056^{+0.021}_{-0.027}$ & $0.036^{+0.014}_{-0.018}$ & $42.7\pm0.2$ \\
    C & $0.030^{+0.031}_{-0.021}$ & $0.018^{+0.018}_{-0.012}$ & $49.0\pm0.2$ \\
    D & $0.072^{+0.034}_{-0.016}$ & $0.073^{+0.035}_{-0.016}$ & $27.5\pm0.2$ \\
\hline
\end{tabular}
\tablefoot{Lensed quasar images are labelled from A to D, $\alpha$ is the stellar mass fraction, $\kappa_{\ast}$ is the convergence in stars, and d is the distance from each quasar image to the BCG.
}
\end{table}

\begin{figure}
    \centering
    \includegraphics[width=\linewidth]{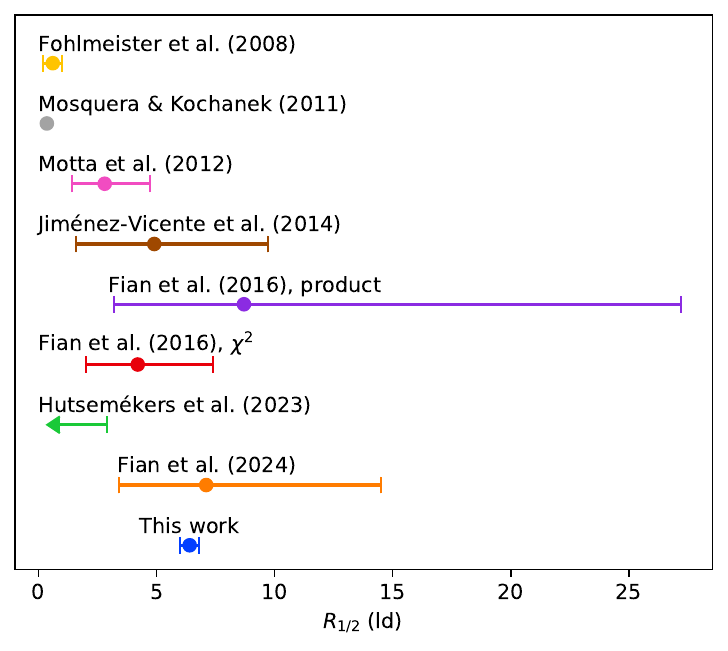}
    \caption{Comparison with previous quasar accretion disk size estimates for SDSS~J1004+4112. The results are reported as the r-band ($\lambda_{\text{obs}}$=6580\AA, $\lambda_{\text{rest}}$=2407\AA) half-light radius in light-days for a mean stellar mass of $0.3M_{\sun}$ to homogenise the comparison.}
    \label{fig:R_half-comp}
\end{figure}

We compare our quasar accretion disk size estimate with previous determinations in Fig. \ref{fig:R_half-comp}. From spectroscopic measurements of images A and B, \citet{2024A&A...682A..57F} inferred a half-light radius of $7.1^{+7.4}_{-3.7}$ light-days for the continuum emission corresponding to the r-band wavelengths, which fully covers our size estimate. Also from spectroscopic data, \citet{2023A&A...672A..45H} reported an upper limit on the continuum disk size of $R_{1/2}<2.9$ light-days for $\langle M \rangle=0.3M_{\sun}$ at 2400\AA, roughly the rest frame centre of the r band, which is significantly smaller than our size estimate. \citet{2016ApJ...830..149F} used light curves of the four quasar images and inferred a half-light radius of $R_{1/2}=8.7^{+18.5}_{-5.5}$ light-days using the histogram product method and $R_{1/2}=4.2^{+3.2}_{-2.2}$ light-days using the corrected $\chi^2$ method. They assumed in both cases a stellar mass of $0.3M_{\sun}$ and that 10\% of the mass is in stars for all images. Their results are totally compatible with our estimate, but we recover tighter constraints because of the longer light curves. \citet{2014ApJ...783...47J} estimated a Bayesian r-band accretion disk size of $R_{1/2}=4.9^{+4.8}_{-3.3}$ light-days for $\langle M \rangle=0.3M_{\sun}$ using a wavelength dependence of $R\propto\lambda^{1.3\pm0.6}$, which is inferred from multi-wavelength microlensing of images A and B. This value is compatible within errors with the size inferred in this work. \citet{2012ApJ...755...82M} inferred a disk size at $\lambda_{\text{rest}}$=3363\AA\ of $r_s=6^{+4}_{-3}$ light-days and a wavelength dependence of $R\propto\lambda^{1.0\pm0.4}$. This implies a half-light radius in the r band of $R_{1/2}=2.8^{+1.9}_{-1.4}$ light-days for $\langle M \rangle=0.3M_{\sun}$, which is in a 1.9$\sigma$ tension with our estimate. We infer significantly larger sizes than \citet{2011ApJ...738...96M} and \citet{2008ApJ...676..761F}, who respectively found $R_{1/2}=0.35$ light-days for a face-on disk in the r band based on the thin-disk scaling relation between flux and size and $R_{1/2}=0.6\pm0.4$ light-days based on the microlensing in images A and B.

The X-ray monitoring of the Fe K$\alpha$ lines in SDSS~J1004+4112 and another lens systems \citep{2012ApJ...755...24C,2017ApJ...837...26C,2019ApJ...879...35D} has enabled the size inference of this emission region via microlensing. \citet{2017ApJ...837...26C} placed an order of magnitude of $\sim20$ gravitational radii ($r_g$) for the emission region by combining three gravitational lens systems. \citet{2019ApJ...879...35D} constrained the Fe K$\alpha$ emission region to $5.9-7.4 r_g$ at the 68\% confidence level for a joint sample of four lenses, including SDSS~J1004+4112. These size estimates for the SDSS~J1004+4112 black hole mass of $3.98\times10^8 M_{\odot}$ \citep{2017ApJ...837...26C} correspond to $\sim0.45$ and $0.13-0.17$ light-days, respectively. Future individual studies of X-ray emission lines and continuum emission in other optical bands will be useful to constrain the structure of this quasar accretion disk. Nonetheless, these X-ray determinations indicate that the source X-ray-emitting region is more compact than the r-band-emitting region, as expected.

It is not straightforward to compare our stellar mass fraction results with other determinations. The stellar mass fraction depends not only on the radial distance from the brightest cluster galaxy (BCG) but also on additional cluster galaxy members along the line of sight where the quasar images are located and on the diffuse matter of the intracluster medium. We can take the baryonic mass fraction of the Universe as a reference for an upper estimate on the mean stellar mass fraction. From \citet{2020A&A...641A...6P} results, the mean fraction of baryons in the Universe is $f_b=\Omega_b/(\Omega_b+\Omega_c)=0.1573\pm0.0013$ ($\Omega_b$ being the baryon density and $\Omega_c$ the cold dark matter density, both in terms of the critical density). Except for $\alpha_A$, where we only obtain a lower limit, the $\alpha$ values are below this upper limit (see Table \ref{tab:stars}). Another upper estimate can be the fraction between the surface density of galaxy cluster members and the total surface density of the cluster at the quasar positions from the \citet{2022ApJ...937...35F} model. These limits are defined as $\alpha^{\text{max}}_i=\sum_n \kappa_{\text{pJaffe},n}(\vec{r_i})/\kappa_{\text{tot}}(\vec{r_i}),$ where $n$ runs over all the modelled galaxies, including the BCG, and $\vec{r_i}$ is the position of the $i$th quasar image. This should be a fair stellar mass fraction limit since the galaxies were modelled as pseudo-Jaffe ellipsoids, which include both the dark matter and baryonic mass in galaxies, and the above definition assumes that all the mass in these galaxies is in stars. The maximum value for each image corresponds to $\alpha^{\text{max}}_A=0.10$, $\alpha^{\text{max}}_B=0.08$, $\alpha^{\text{max}}_C=0.03,$ and $\alpha^{\text{max}}_D=0.12$. The stellar mass fractions' lower confidence levels are at least 30\% of the $\alpha^{\text{max}}_i$ estimates. Even though the central values do not surpass the mass fraction in pseudo-Jaffe components, these percentages are rather high because one would expect, on average, $\sim$20\% of the mass in galaxies to be contained in stars \citep{2015ApJ...799..149J}. Hence, it is likely that stars in the intracluster medium are contributing to increase the total stellar mass fraction.

A lower bound could be the mean stellar mass fraction of entire galaxy clusters given that in this system the quasar images are located within R$<$50 kpc/h of the centre. Following the stellar mass fraction relation of \citet{2009ApJ...703..982G} obtained from observed galaxy groups and poor clusters, we find $f_{500}^{\text{stars}}=0.014\pm0.002$ for a $M_{500}=9.89\times10^{14} M_{\sun}/h$ \citep[derived from the mass model of][]{2022ApJ...937...35F}. \citet{2013ApJ...778...14G} derived a similar relation for the projected stellar mass with respect to $M_{500}$. Dividing this relation by $M_{500}$, we find a mean stellar mass fraction of $0.0119\pm0.0014$ for our cluster mass. \citet{2018AstL...44....8K} added more massive clusters to the \citet{2013ApJ...778...14G} sample and derived a $M_{\ast,\text{tot}}-M_{500}$ that resulted in a stellar mass fraction of $0.0088\pm0.0012$. From TNG100 and TNG300\footnote{The Next Generation Illustris simulations with side lengths of $\sim$100 Mpc and $\sim$300 Mpc, respectively.} galaxy group and cluster simulations, \citet{2018MNRAS.475..648P} derived the same type of relation as above, obtaining a stellar mass fraction of around 0.0149. All these observational or numerical works infer a mean stellar mass fraction of around 1\% for entire galaxy clusters. This value is within the confidence interval estimated for $\alpha_C$, but for $\alpha_A$, $\alpha_B$, and $\alpha_D$ our estimations are larger.

On the other hand, the radial dependence of the stellar mass due to the BCG and the intracluster light (ICL) offers a minimum stellar mass estimate at the specific image positions. They are usually reported as the enclosed stellar mass within a specified radius. This can be translated to a stellar surface density, and, when divided by the critical surface density of the lens system $\Sigma_{\text{crit}}=3.23\times 10^9\ \text{h}\ M_{\sun}\ \text{kpc}^{-2}$ and by the total convergence at the quasar positions, an estimate of $\alpha$ for each quasar image can be obtained. \citet{2018AstL...44....8K} reported the enclosed stellar mass of the BCG plus the ICL in the inner 30, 50, and 70 kpc for nine observed galaxy clusters. \citet{2018MNRAS.474.3009D} also inferred the stellar masses within 10, 50, and 100 kpc for 23 galaxy clusters; for 16 of them they found $M_{500}>10^{14} M_{\sun}$, which is more similar to SDSS~J1004+4112. \citet{2020MNRAS.498.2114H} reported the radial stellar surface density profile of galaxy clusters from \textsc{fable}\footnote{Feedback Acting on Baryons in Large-scale Environments cosmological hydrodynamical simulation.} simulations. The resulting mean stellar mass fractions and the standard deviations at the locations of each quasar image from these works are reported in Table \ref{tab:ICL-est}.

\begin{table}
\caption{Estimates based on previous works for the stellar mass fraction.}
\label{tab:ICL-est}
\centering
\begin{tabular}{c c c c}
\hline\hline
BCG+ICL & K18 & D18 & H20 \\
\hline
    $\alpha_A$ & $0.012\pm0.007$ & $0.006\pm0.002$ & $0.012^{+0.022}_{-0.008}$ \\
    $\alpha_B$ & $0.013\pm0.008$ & $0.007\pm0.003$ & $0.013^{+0.025}_{-0.009}$ \\
    $\alpha_C$ & $0.014\pm0.009$ & $0.008\pm0.003$ & $0.011^{+0.022}_{-0.007}$ \\
    $\alpha_D$ & $0.018\pm0.010$ & $0.038\pm0.013$ & $0.022^{+0.028}_{-0.012}$ \\
\hline
\end{tabular}
\tablefoot{Estimated contribution to the stellar mass fraction from the BCG and the ICL at the quasar image positions from the works of K18 \citep{2018AstL...44....8K}, D18 \citep{2018MNRAS.474.3009D}, and H20 \citep{2020MNRAS.498.2114H}.
}
\end{table}

These BCG+ICL estimates are lower limits because the contributions from the galaxy cluster members are excluded. To obtain more realistic estimates, the 20\% of the mass fraction from galaxy members (excluding the BCG) in the \citet{2022ApJ...937...35F} model ($0.008$, $0.001$, $0.002,$ and $0.008$, for images A, B, C, and D, respectively) can be added to the values estimated by \citet{2018AstL...44....8K}, \citet{2018MNRAS.474.3009D}, and \citet{2020MNRAS.498.2114H}. The average discrepancies for the stellar mass fractions at the positions of images A, B, C, and D are $2.16$, $1.47$, $0.69,$ and $1.65\sigma$, respectively. For the discrepancy of image A, we used its central value, 0.090 (where the minimum $\chi^2$ is found) and the lower error, 0.031 (the difference between the central value and the lower estimate). Hence, the stellar mass fraction at image C is completely consistent with the estimates, at images B and D a slight tension is found, and at image A our inference is 2$\sigma$ discrepant with these estimates. The \citet{2008ApJ...676..761F} models also preferred larger values of $\kappa_{\ast}$ for images A and B, and the authors suggested that the microlensing was probably due to a satellite galaxy rather than stars in the ICL. Recently, \citet{2024MNRAS.527.2639P} built free-form and hybrid mass models for SDSS~J1004+4112 and found in all models a mass clump of around 2\arcsec\ located at $\sim$(5,-1)\arcsec\ from quasar image A. The nature and properties of this possible clump are still undefined, but it remains present after the matter is redistributed following the procedure described in \citet{2024MNRAS.529.1222L}. Although this substructure may not be responsible for the slightly larger stellar mass fractions that we estimated, this finding indicates that the mass structure of SDSS~J1004+4112 may be complicated and composed of substructures yet to be detected. Additional stellar components may help explain the inferred stellar mass fractions being higher than the fractions expected from galaxy members and from ICL studies, but observational evidence should be provided to confirm this hypothesis. If microlensing is being produced by an undetected individual galaxy member, it must be aligned with the line of sight of images and, hence, the quasar emission can mask it, making it difficult to detect. On the other hand, the stellar mass fraction can be explained if diffuse components (mainly in the form of stars) are present in these regions. The origin of this component could be a substructure in the cluster galaxy or a stellar tidal stream from the infall of satellite galaxies into the BCG.

In this work we attempted to estimate the microlensing parameters using magnification fluctuation statistics. This kind of microlensing study in which the time dependence is not included seems to obtain results consistent with other techniques while keeping the computational cost low \citep[see][]{2016ApJ...830..149F,2018ApJ...869..132F,2021A&A...654A..70F}. Nonetheless, the temporally correlated microlensing, especially the event detected between days 6500 and 7500 in the A$-$B and A$-$C difference light curves of Fig. \ref{fig:micro-diff}, may be worthy of study in a future work with direct light curve fitting \citep{2004ApJ...605...58K} to further test the validity of the method employed in this work.

In conclusion, we have determined the quasar accretion disk size and the stellar mass fraction at the four brightest quasar image positions in SDSS~J1004+4112 using the 14.5-year light curves and time delays reported by \citet{2022ApJ...937...34M}. We were able to constrain, for the first time, the stellar mass fraction at the individual image positions. Our quasar disk size is broadly compatible with previous works, and the stellar mass fractions we find lie between the expected upper and lower bounds extracted from the literature. The stellar mass fraction at the position of image C can be mostly explained by the contributions of the BCG, the ICL, and the known galaxy cluster members. At images B and D, the inferred mass fractions are slightly above expectations, with discrepancies around 1.5$\sigma$. For image A, only a lower limit has been obtained, and the central value is in a 2$\sigma$ tension with the contribution of the known stellar components. These inferred stellar mass fractions suggest that other stellar components are present in the regions where images B, D, and especially image A are located.
 
\begin{acknowledgements}

We thank C. S. Kochanek for his useful comments and suggestions throughout this work. We are also grateful for the anonymous referee's reviews, which have improved the final version of the article. This research was supported by the grants PID2020-118687GB-C31, PID2020-118687GB-C32 and PID2020-118687GB-C33 financed by the Spanish Ministerio de Ciencia e Innovaci\'on through MCIN/AEI/10.13039/501100011033. J.A.M. is also supported by the project of excellence PROMETEO CIPROM/2023/21 of the Conselleria de Educación, Universidades y Empleo (Generalitat Valenciana). J.J.V. is also supported by projects FQM-108, P20\_00334, and A-FQM-510-UGR20/FEDER, financed by Junta de Andaluc\'{\i}a.
\end{acknowledgements}

\bibliographystyle{aa}
\bibliography{paper}

\begin{thebibliography}{53}
\expandafter\ifx\csname natexlab\endcsname\relax\def\natexlab#1{#1}\fi

\bibitem[{{Abajas} {et~al.}(2007){Abajas}, {Mediavilla}, {Mu{\~n}oz},
  {G{\'o}mez-{\'A}lvarez}, \& {Gil-Merino}}]{2007ApJ...658..748A}
{Abajas}, C., {Mediavilla}, E., {Mu{\~n}oz}, J.~A., {G{\'o}mez-{\'A}lvarez},
  P., \& {Gil-Merino}, R. 2007, \apj, 658, 748

\bibitem[{{Awad} {et~al.}(2023){Awad}, {Chan}, {Millon}, {Courbin}, \&
  {Paic}}]{2023A&A...673A..88A}
{Awad}, P., {Chan}, J.~H.~H., {Millon}, M., {Courbin}, F., \& {Paic}, E. 2023,
  \aap, 673, A88

\bibitem[{{Blackburne} {et~al.}(2015){Blackburne}, {Kochanek}, {Chen}, {Dai},
  \& {Chartas}}]{2015ApJ...798...95B}
{Blackburne}, J.~A., {Kochanek}, C.~S., {Chen}, B., {Dai}, X., \& {Chartas}, G.
  2015, \apj, 798, 95

\bibitem[{{Chang} \& {Refsdal}(1979)}]{1979Natur.282..561C}
{Chang}, K. \& {Refsdal}, S. 1979, \nat, 282, 561

\bibitem[{{Chartas} {et~al.}(2017){Chartas}, {Krawczynski}, {Zalesky},
  {Kochanek}, {Dai}, {Morgan}, \& {Mosquera}}]{2017ApJ...837...26C}
{Chartas}, G., {Krawczynski}, H., {Zalesky}, L., {et~al.} 2017, \apj, 837, 26

\bibitem[{{Chen} {et~al.}(2012){Chen}, {Dai}, {Kochanek}, {Chartas},
  {Blackburne}, \& {Morgan}}]{2012ApJ...755...24C}
{Chen}, B., {Dai}, X., {Kochanek}, C.~S., {et~al.} 2012, \apj, 755, 24

\bibitem[{{Dai} {et~al.}(2019){Dai}, {Steele}, {Guerras}, {Morgan}, \&
  {Chen}}]{2019ApJ...879...35D}
{Dai}, X., {Steele}, S., {Guerras}, E., {Morgan}, C.~W., \& {Chen}, B. 2019,
  \apj, 879, 35

\bibitem[{{DeMaio} {et~al.}(2018){DeMaio}, {Gonzalez}, {Zabludoff}, {Zaritsky},
  {Connor}, {Donahue}, \& {Mulchaey}}]{2018MNRAS.474.3009D}
{DeMaio}, T., {Gonzalez}, A.~H., {Zabludoff}, A., {et~al.} 2018, \mnras, 474,
  3009

\bibitem[{{Fian} {et~al.}(2018{\natexlab{a}}){Fian}, {Guerras}, {Mediavilla},
  {Jim{\'e}nez-Vicente}, {Mu{\~n}oz}, {Falco}, {Motta}, \&
  {Hanslmeier}}]{2018ApJ...859...50F}
{Fian}, C., {Guerras}, E., {Mediavilla}, E., {et~al.} 2018{\natexlab{a}}, \apj,
  859, 50

\bibitem[{{Fian} {et~al.}(2016){Fian}, {Mediavilla}, {Hanslmeier}, {Oscoz},
  {Serra-Ricart}, {Mu{\~n}oz}, \& {Jim{\'e}nez-Vicente}}]{2016ApJ...830..149F}
{Fian}, C., {Mediavilla}, E., {Hanslmeier}, A., {et~al.} 2016, \apj, 830, 149

\bibitem[{{Fian} {et~al.}(2021{\natexlab{a}}){Fian}, {Mediavilla},
  {Jim{\'e}nez-Vicente}, {Motta}, {Mu{\~n}oz}, {Chelouche},
  {Gom{\'e}z-Alvarez}, {Rojas}, \& {Hanslmeier}}]{2021A&A...654A..70F}
{Fian}, C., {Mediavilla}, E., {Jim{\'e}nez-Vicente}, J., {et~al.}
  2021{\natexlab{a}}, \aap, 654, A70

\bibitem[{{Fian} {et~al.}(2018{\natexlab{b}}){Fian}, {Mediavilla},
  {Jim{\'e}nez-Vicente}, {Mu{\~n}oz}, \& {Hanslmeier}}]{2018ApJ...869..132F}
{Fian}, C., {Mediavilla}, E., {Jim{\'e}nez-Vicente}, J., {Mu{\~n}oz}, J.~A., \&
  {Hanslmeier}, A. 2018{\natexlab{b}}, \apj, 869, 132

\bibitem[{{Fian} {et~al.}(2021{\natexlab{b}}){Fian}, {Mediavilla}, {Motta},
  {Jim{\'e}nez-Vicente}, {Mu{\~n}oz}, {Chelouche}, \&
  {Hanslmeier}}]{2021A&A...653A.109F}
{Fian}, C., {Mediavilla}, E., {Motta}, V., {et~al.} 2021{\natexlab{b}}, \aap,
  653, A109

\bibitem[{{Fian} {et~al.}(2024){Fian}, {Mu{\~n}oz}, {For{\'e}s-Toribio},
  {Mediavilla}, {Jim{\'e}nez-Vicente}, {Chelouche}, {Kaspi}, \&
  {Richards}}]{2024A&A...682A..57F}
{Fian}, C., {Mu{\~n}oz}, J.~A., {For{\'e}s-Toribio}, R., {et~al.} 2024, \aap,
  682, A57

\bibitem[{{Fohlmeister} {et~al.}(2008){Fohlmeister}, {Kochanek}, {Falco},
  {Morgan}, \& {Wambsganss}}]{2008ApJ...676..761F}
{Fohlmeister}, J., {Kochanek}, C.~S., {Falco}, E.~E., {Morgan}, C.~W., \&
  {Wambsganss}, J. 2008, \apj, 676, 761

\bibitem[{{For{\'e}s-Toribio} {et~al.}(2022){For{\'e}s-Toribio}, {Mu{\~n}oz},
  {Kochanek}, \& {Mediavilla}}]{2022ApJ...937...35F}
{For{\'e}s-Toribio}, R., {Mu{\~n}oz}, J.~A., {Kochanek}, C.~S., \&
  {Mediavilla}, E. 2022, \apj, 937, 35

\bibitem[{{Giodini} {et~al.}(2009){Giodini}, {Pierini}, {Finoguenov}, {Pratt},
  {Boehringer}, {Leauthaud}, {Guzzo}, {Aussel}, {Bolzonella}, {Capak}, {Elvis},
  {Hasinger}, {Ilbert}, {Kartaltepe}, {Koekemoer}, {Lilly}, {Massey},
  {McCracken}, {Rhodes}, {Salvato}, {Sanders}, {Scoville}, {Sasaki}, {Smolcic},
  {Taniguchi}, {Thompson}, \& {COSMOS Collaboration}}]{2009ApJ...703..982G}
{Giodini}, S., {Pierini}, D., {Finoguenov}, A., {et~al.} 2009, \apj, 703, 982

\bibitem[{{G{\'o}mez-{\'A}lvarez} {et~al.}(2006){G{\'o}mez-{\'A}lvarez},
  {Mediavilla}, {Mu{\~n}oz}, {Arribas}, {S{\'a}nchez}, {Oscoz}, {Prada}, \&
  {Serra-Ricart}}]{2006ApJ...645L...5G}
{G{\'o}mez-{\'A}lvarez}, P., {Mediavilla}, E., {Mu{\~n}oz}, J.~A., {et~al.}
  2006, \apjl, 645, L5

\bibitem[{{Gonzalez} {et~al.}(2013){Gonzalez}, {Sivanandam}, {Zabludoff}, \&
  {Zaritsky}}]{2013ApJ...778...14G}
{Gonzalez}, A.~H., {Sivanandam}, S., {Zabludoff}, A.~I., \& {Zaritsky}, D.
  2013, \apj, 778, 14

\bibitem[{{Green}(2006)}]{2006ApJ...644..733G}
{Green}, P.~J. 2006, \apj, 644, 733

\bibitem[{{Hainline} {et~al.}(2013){Hainline}, {Morgan}, {MacLeod}, {Landaal},
  {Kochanek}, {Harris}, {Tilleman}, {Goicoechea}, {Shalyapin}, \&
  {Falco}}]{2013ApJ...774...69H}
{Hainline}, L.~J., {Morgan}, C.~W., {MacLeod}, C.~L., {et~al.} 2013, \apj, 774,
  69

\bibitem[{{Hartley} {et~al.}(2021){Hartley}, {Jackson}, {Badole}, {McKean},
  {Sluse}, \& {Vives-Arias}}]{2021MNRAS.508.4625H}
{Hartley}, P., {Jackson}, N., {Badole}, S., {et~al.} 2021, \mnras, 508, 4625

\bibitem[{{Henden} {et~al.}(2020){Henden}, {Puchwein}, \&
  {Sijacki}}]{2020MNRAS.498.2114H}
{Henden}, N.~A., {Puchwein}, E., \& {Sijacki}, D. 2020, \mnras, 498, 2114

\bibitem[{{Hutsem{\'e}kers} {et~al.}(2023){Hutsem{\'e}kers}, {Sluse},
  {Savi{\'c}}, \& {Richards}}]{2023A&A...672A..45H}
{Hutsem{\'e}kers}, D., {Sluse}, D., {Savi{\'c}}, {\DJ}., \& {Richards}, G.~T.
  2023, \aap, 672, A45

\bibitem[{{Inada} {et~al.}(2003){Inada}, {Oguri}, {Pindor}, {Hennawi}, {Chiu},
  {Zheng}, {Ichikawa}, {Gregg}, {Becker}, {Suto}, {Strauss}, {Turner},
  {Keeton}, {Annis}, {Castander}, {Eisenstein}, {Frieman}, {Fukugita}, {Gunn},
  {Johnston}, {Kent}, {Nichol}, {Richards}, {Rix}, {Sheldon}, {Bahcall},
  {Brinkmann}, {Ivezi{\'c}}, {Lamb}, {McKay}, {Schneider}, \&
  {York}}]{2003Natur.426..810I}
{Inada}, N., {Oguri}, M., {Pindor}, B., {et~al.} 2003, \nat, 426, 810

\bibitem[{{Jackson}(2020)}]{2020ApJ...900L..15J}
{Jackson}, N. 2020, \apjl, 900, L15

\bibitem[{{James}(2006)}]{2006smep.book.....J}
{James}, F. 2006, {Statistical Methods in Experimental Physics: 2nd Edition}
  ({World Scientific Publishing Co., Pte. Ltd.})

\bibitem[{{Jim{\'e}nez-Vicente} \& {Mediavilla}(2022)}]{2022ApJ...941...80J}
{Jim{\'e}nez-Vicente}, J. \& {Mediavilla}, E. 2022, \apj, 941, 80

\bibitem[{{Jim{\'e}nez-Vicente} {et~al.}(2015){Jim{\'e}nez-Vicente},
  {Mediavilla}, {Kochanek}, \& {Mu{\~n}oz}}]{2015ApJ...799..149J}
{Jim{\'e}nez-Vicente}, J., {Mediavilla}, E., {Kochanek}, C.~S., \& {Mu{\~n}oz},
  J.~A. 2015, \apj, 799, 149

\bibitem[{{Jim{\'e}nez-Vicente} {et~al.}(2014){Jim{\'e}nez-Vicente},
  {Mediavilla}, {Kochanek}, {Mu{\~n}oz}, {Motta}, {Falco}, \&
  {Mosquera}}]{2014ApJ...783...47J}
{Jim{\'e}nez-Vicente}, J., {Mediavilla}, E., {Kochanek}, C.~S., {et~al.} 2014,
  \apj, 783, 47

\bibitem[{{Jim{\'e}nez-Vicente} {et~al.}(2012){Jim{\'e}nez-Vicente},
  {Mediavilla}, {Mu{\~n}oz}, \& {Kochanek}}]{2012ApJ...751..106J}
{Jim{\'e}nez-Vicente}, J., {Mediavilla}, E., {Mu{\~n}oz}, J.~A., \& {Kochanek},
  C.~S. 2012, \apj, 751, 106

\bibitem[{{Kochanek}(2004)}]{2004ApJ...605...58K}
{Kochanek}, C.~S. 2004, \apj, 605, 58

\bibitem[{{Kravtsov} {et~al.}(2018){Kravtsov}, {Vikhlinin}, \&
  {Meshcheryakov}}]{2018AstL...44....8K}
{Kravtsov}, A.~V., {Vikhlinin}, A.~A., \& {Meshcheryakov}, A.~V. 2018,
  Astronomy Letters, 44, 8

\bibitem[{{Lamer} {et~al.}(2006){Lamer}, {Schwope}, {Wisotzki}, \&
  {Christensen}}]{2006A&A...454..493L}
{Lamer}, G., {Schwope}, A., {Wisotzki}, L., \& {Christensen}, L. 2006, \aap,
  454, 493

\bibitem[{{Liesenborgs} {et~al.}(2024){Liesenborgs}, {Perera}, \&
  {Williams}}]{2024MNRAS.529.1222L}
{Liesenborgs}, J., {Perera}, D., \& {Williams}, L. L.~R. 2024, \mnras, 529,
  1222

\bibitem[{{MacLeod} {et~al.}(2015){MacLeod}, {Morgan}, {Mosquera}, {Kochanek},
  {Tewes}, {Courbin}, {Meylan}, {Chen}, {Dai}, \&
  {Chartas}}]{2015ApJ...806..258M}
{MacLeod}, C.~L., {Morgan}, C.~W., {Mosquera}, A., {et~al.} 2015, \apj, 806,
  258

\bibitem[{{McKean} {et~al.}(2021){McKean}, {Luichies}, {Drabent}, {G{\"u}rkan},
  {Hartley}, {Lafontaine}, {Prandoni}, {R{\"o}ttgering}, {Shimwell}, {Stacey},
  \& {Tasse}}]{2021MNRAS.505L..36M}
{McKean}, J.~P., {Luichies}, R., {Drabent}, A., {et~al.} 2021, \mnras, 505, L36

\bibitem[{{Mediavilla} {et~al.}(2009){Mediavilla}, {Mu{\~n}oz}, {Falco},
  {Motta}, {Guerras}, {Canovas}, {Jean}, {Oscoz}, \&
  {Mosquera}}]{2009ApJ...706.1451M}
{Mediavilla}, E., {Mu{\~n}oz}, J.~A., {Falco}, E., {et~al.} 2009, \apj, 706,
  1451

\bibitem[{{Morgan} {et~al.}(2018){Morgan}, {Hyer}, {Bonvin}, {Mosquera},
  {Cornachione}, {Courbin}, {Kochanek}, \& {Falco}}]{2018ApJ...869..106M}
{Morgan}, C.~W., {Hyer}, G.~E., {Bonvin}, V., {et~al.} 2018, \apj, 869, 106

\bibitem[{{Mortonson} {et~al.}(2005){Mortonson}, {Schechter}, \&
  {Wambsganss}}]{2005ApJ...628..594M}
{Mortonson}, M.~J., {Schechter}, P.~L., \& {Wambsganss}, J. 2005, \apj, 628,
  594

\bibitem[{{Mosquera} \& {Kochanek}(2011)}]{2011ApJ...738...96M}
{Mosquera}, A.~M. \& {Kochanek}, C.~S. 2011, \apj, 738, 96

\bibitem[{{Motta} {et~al.}(2012){Motta}, {Mediavilla}, {Falco}, \&
  {Mu{\~n}oz}}]{2012ApJ...755...82M}
{Motta}, V., {Mediavilla}, E., {Falco}, E., \& {Mu{\~n}oz}, J.~A. 2012, \apj,
  755, 82

\bibitem[{{Mu{\~n}oz} {et~al.}(2022){Mu{\~n}oz}, {Kochanek}, {Fohlmeister},
  {Wambsganss}, {Falco}, \& {For{\'e}s-Toribio}}]{2022ApJ...937...34M}
{Mu{\~n}oz}, J.~A., {Kochanek}, C.~S., {Fohlmeister}, J., {et~al.} 2022, \apj,
  937, 34

\bibitem[{{Mu{\~n}oz} {et~al.}(2016){Mu{\~n}oz}, {Vives-Arias}, {Mosquera},
  {Jim{\'e}nez-Vicente}, {Kochanek}, \& {Mediavilla}}]{2016ApJ...817..155M}
{Mu{\~n}oz}, J.~A., {Vives-Arias}, H., {Mosquera}, A.~M., {et~al.} 2016, \apj,
  817, 155

\bibitem[{{Oguri}(2010)}]{2010PASJ...62.1017O}
{Oguri}, M. 2010, \pasj, 62, 1017

\bibitem[{{Perera} {et~al.}(2024){Perera}, {Williams}, {Liesenborgs}, {Ghosh},
  \& {Saha}}]{2024MNRAS.527.2639P}
{Perera}, D., {Williams}, L. L.~R., {Liesenborgs}, J., {Ghosh}, A., \& {Saha},
  P. 2024, \mnras, 527, 2639

\bibitem[{{Pillepich} {et~al.}(2018){Pillepich}, {Nelson}, {Hernquist},
  {Springel}, {Pakmor}, {Torrey}, {Weinberger}, {Genel}, {Naiman}, {Marinacci},
  \& {Vogelsberger}}]{2018MNRAS.475..648P}
{Pillepich}, A., {Nelson}, D., {Hernquist}, L., {et~al.} 2018, \mnras, 475, 648

\bibitem[{{Planck Collaboration} {et~al.}(2020){Planck Collaboration},
  {Aghanim}, {Akrami}, {Ashdown}, {Aumont}, {Baccigalupi}, {Ballardini},
  {Banday}, {Barreiro}, {Bartolo}, {Basak}, {Battye}, {Benabed}, {Bernard},
  {Bersanelli}, {Bielewicz}, {Bock}, {Bond}, {Borrill}, {Bouchet}, {Boulanger},
  {Bucher}, {Burigana}, {Butler}, {Calabrese}, {Cardoso}, {Carron},
  {Challinor}, {Chiang}, {Chluba}, {Colombo}, {Combet}, {Contreras}, {Crill},
  {Cuttaia}, {de Bernardis}, {de Zotti}, {Delabrouille}, {Delouis}, {Di
  Valentino}, {Diego}, {Dor{\'e}}, {Douspis}, {Ducout}, {Dupac}, {Dusini},
  {Efstathiou}, {Elsner}, {En{\ss}lin}, {Eriksen}, {Fantaye}, {Farhang},
  {Fergusson}, {Fernandez-Cobos}, {Finelli}, {Forastieri}, {Frailis},
  {Fraisse}, {Franceschi}, {Frolov}, {Galeotta}, {Galli}, {Ganga},
  {G{\'e}nova-Santos}, {Gerbino}, {Ghosh}, {Gonz{\'a}lez-Nuevo}, {G{\'o}rski},
  {Gratton}, {Gruppuso}, {Gudmundsson}, {Hamann}, {Handley}, {Hansen},
  {Herranz}, {Hildebrandt}, {Hivon}, {Huang}, {Jaffe}, {Jones}, {Karakci},
  {Keih{\"a}nen}, {Keskitalo}, {Kiiveri}, {Kim}, {Kisner}, {Knox},
  {Krachmalnicoff}, {Kunz}, {Kurki-Suonio}, {Lagache}, {Lamarre}, {Lasenby},
  {Lattanzi}, {Lawrence}, {Le Jeune}, {Lemos}, {Lesgourgues}, {Levrier},
  {Lewis}, {Liguori}, {Lilje}, {Lilley}, {Lindholm}, {L{\'o}pez-Caniego},
  {Lubin}, {Ma}, {Mac{\'\i}as-P{\'e}rez}, {Maggio}, {Maino}, {Mandolesi},
  {Mangilli}, {Marcos-Caballero}, {Maris}, {Martin}, {Martinelli},
  {Mart{\'\i}nez-Gonz{\'a}lez}, {Matarrese}, {Mauri}, {McEwen}, {Meinhold},
  {Melchiorri}, {Mennella}, {Migliaccio}, {Millea}, {Mitra},
  {Miville-Desch{\^e}nes}, {Molinari}, {Montier}, {Morgante}, {Moss}, {Natoli},
  {N{\o}rgaard-Nielsen}, {Pagano}, {Paoletti}, {Partridge}, {Patanchon},
  {Peiris}, {Perrotta}, {Pettorino}, {Piacentini}, {Polastri}, {Polenta},
  {Puget}, {Rachen}, {Reinecke}, {Remazeilles}, {Renzi}, {Rocha}, {Rosset},
  {Roudier}, {Rubi{\~n}o-Mart{\'\i}n}, {Ruiz-Granados}, {Salvati}, {Sandri},
  {Savelainen}, {Scott}, {Shellard}, {Sirignano}, {Sirri}, {Spencer},
  {Sunyaev}, {Suur-Uski}, {Tauber}, {Tavagnacco}, {Tenti}, {Toffolatti},
  {Tomasi}, {Trombetti}, {Valenziano}, {Valiviita}, {Van Tent}, {Vibert},
  {Vielva}, {Villa}, {Vittorio}, {Wandelt}, {Wehus}, {White}, {White},
  {Zacchei}, \& {Zonca}}]{2020A&A...641A...6P}
{Planck Collaboration}, {Aghanim}, N., {Akrami}, Y., {et~al.} 2020, \aap, 641,
  A6

\bibitem[{{Poindexter} \& {Kochanek}(2010)}]{2010ApJ...712..658P}
{Poindexter}, S. \& {Kochanek}, C.~S. 2010, \apj, 712, 658

\bibitem[{{Popovi{\'c}} {et~al.}(2020){Popovi{\'c}}, {Afanasiev}, {Moiseev},
  {Smirnova}, {Simi{\'c}}, {Savi{\'c}}, {Mediavilla}, \&
  {Fian}}]{2020A&A...634A..27P}
{Popovi{\'c}}, L.~{\v{C}}., {Afanasiev}, V.~L., {Moiseev}, A., {et~al.} 2020,
  \aap, 634, A27

\bibitem[{{Richards} {et~al.}(2004){Richards}, {Keeton}, {Pindor}, {Hennawi},
  {Hall}, {Turner}, {Inada}, {Oguri}, {Ichikawa}, {Becker}, {Gregg}, {White},
  {Wyithe}, {Schneider}, {Johnston}, {Frieman}, \&
  {Brinkmann}}]{2004ApJ...610..679R}
{Richards}, G.~T., {Keeton}, C.~R., {Pindor}, B., {et~al.} 2004, \apj, 610, 679

\bibitem[{{Ross} {et~al.}(2009){Ross}, {Assef}, {Kochanek}, {Falco}, \&
  {Poindexter}}]{2009ApJ...702..472R}
{Ross}, N.~R., {Assef}, R.~J., {Kochanek}, C.~S., {Falco}, E., \& {Poindexter},
  S.~D. 2009, \apj, 702, 472

\bibitem[{{Vernardos} \& {Tsagkatakis}(2019)}]{2019MNRAS.486.1944V}
{Vernardos}, G. \& {Tsagkatakis}, G. 2019, \mnras, 486, 1944

\end{thebibliography}

\end{document}